# From Bit To Bedside: A Practical Framework For Artificial Intelligence Product Development In Healthcare

David Higgins[1] and Vince I. Madai[2]

[1]Berlin Institute of Health, Berlin, Germany
[2]Charite Lab for Artificial Intelligence in Medicine - CLAIM, Charité Unversitätsmedizin Berlin, Berlin, Germany

## Corresponding Author Details

Dr. David Higgins
Berlin Institute of Health, Anna-Louisa-Karchstr. 2, 10178 Berlin, Germany
E-mail: dave@uiginn.com

Dr. Vince I. Madai
Charité – Universitätsmedizin Berlin, Augustenburger Platz 1,13353 Berlin, Germany
E-mail: vince.madai@gmail.com

# Abstract

Artificial Intelligence (AI) in healthcare holds great potential to expand access to high-quality medical care, whilst reducing overall systemic costs. Despite hitting the headlines regularly and many publications of proofs-of-concept, certified products are failing to breakthrough to the clinic. AI in healthcare is a multi-party process with deep knowledge required in multiple individual domains. The lack of understanding of the specific challenges in the domain is, therefore, the major contributor to the failure to deliver on the big promises. Thus, we present a 'decision perspective' framework, for the development of AI-driven biomedical products, from conception to market launch. Our framework highlights the risks, objectives and key results which are typically required to proceed through a three-phase process to the market launch of a validated medical AI product. We focus on issues related to Clinical validation, Regulatory affairs, Data strategy and Algorithmic development. The development process we propose for AI in healthcare software strongly diverges from modern consumer software development processes. We highlight the key time points to guide founders, investors and key stakeholders throughout their relevant part of the process. Our framework should be seen as a template for innovation frameworks, which can be used to coordinate team communications and responsibilities towards a reasonable product development roadmap, thus unlocking the potential of AI in medicine.

# Introduction

Healthcare systems all over the world face tremendous challenges. The age-related illness burden is increasing, particularly in wealthy countries, due to ageing populations. Lifetime risk for cancer has reached up to 50% [1] and lifetime risk for stroke is 25% [2]. As a consequence of the increasing incidence of these costly diseases, along with broadening access to healthcare and major advances in pharmaceutical and technological disease treatment, costs are increasing enormously. Thus, there is an urgent need to deliver better quality healthcare while simultaneously lowering costs in order to keep our health care systems sustainable.

Technological innovation appears to be the only mechanism which has the potential to fulfil both of these, seemingly contradictory, requirements. Artificial Intelligence (AI), more specifically machine learning(ML)-based AI, has become one of the main technologies driving the so-called 4th industrial revolution [3]. The main ML method at the forefront are so-called deep artificial neural networks (ANNs). ANNs are inspired by simplified neuronal structures and consist of layers of artificial neurons. The strengths of the connection between neurons of different layers are determined in the training process and also determine the predictive capability of the model [4]. ANNs are general function approximators, in theory, capable of approximating any mathematical function of data [5]). Other non-ANN machine learning methods successfully applied in healthcare include decision tree algorithms [6–8], generalised linear models [9], support vector machines [10,11] and gaussian processes (GPs) [12,13]. Modern ML methods constitute the new state-of-the-art in computer vision [14], natural language processing [15] and recommender systems [16–18] facilitating technologies from smart-assistants to self-driving cars. Increasingly, they are applied for healthcare use cases. They have the potential to deliver personalised treatments [19–21] and monitoring [22,23], with lower error rates [23,24], at greatly reduced costs [20,23].

Many publications share the vision that AI in healthcare will achieve the above-mentioned goals to keep our healthcare systems sustainable, e.g. [19,20,25–27]. However, the overwhelming majority of reported positive studies to date are retrospective in nature and the presented solutions, more often than not, provide mere proofs-of-concept as evidenced by systematic reviews [28,29]. AI/ML-based tools which are applicable in the clinical setting - with medical device certification, clinical validation and routine clinical use - are still scarce. Despite a rapidly increasing number of scientific publications describing potential applications, products benefiting patients and our healthcare systems are not deployed at the rates necessary. This *translational*

*gap* constitutes a major public health challenge. One of the most promising technologies of recent years is not reaching the patients and healthcare systems in need.

As the translational gap refers to a block on transforming proofs-of-concept into actual products, we are facing a crisis of AI in healthcare product development. Thus, in the present work, we outline a novel framework of AI in healthcare product development. Product development is defined as "the transformation of a market opportunity and a set of assumptions about product technology into a product available for sale" [30]. It is possible to focus on product development from different angles, i.e. marketing, organisational structure, engineering and operations [30]. However, these views are highly specialised and have a tendency to over- or under-focus on certain challenges. We base our framework on the model of the "decision perspective" where a string of decisions transform an idea into a product [30]. This model is based on the observation that "what is being decided" is fairly consistent over the course of all types of product development, whereas how decisions are implemented can vary tremendously [30]. Thus, importantly, our framework does not give many recommendations on *how* to implement the decisions that are suggested. We focus largely on *what* needs to be decided for a certain stage in development to proceed successfully to the next stage and what needs to be decided to keep the risks of failure for further stages as low as possible.

The main finding that underlies the presented product development framework is that, in contrast to other software products, healthcare-specific differences exist in the development path of AI healthcare products. Product development for a typical technology start-up is long and can be broken down into multiple stages [31]. Typically there is an initial *Formation stage*, in which a team comes together around a potential product idea. This is followed by a *Validation stage*, in which prototypes are turned into products and an eventual business model is settled-on. And, finally, there is a *Scaling phase*, where the addition of considerable capital allows the small company, with a now validated product-market fit, to quickly expand to capture considerable market share of an existing market or to expand into a newly discovered market niche. In this approach, current methodologies drive entrepreneurs towards early contact with potential customers and early trials of incomplete versions of the final product [32]. For AI products aimed at healthcare, the main hidden difficulty is that in order to capture the enhanced value available for medical treatments evidence must first be presented that the product is safe. Additionally, to be incorporated into standards of care, the product must also be shown to either enhance treatment outcomes or decrease costs. This requires a validation pipeline tailored to regulatory and clinical requirements [33], which makes data acquisition and analysis time-consuming and difficult. Current software product development practices are rarely a fit for both the medical certification as well as the clinical validation process. This - in turn - leads to a) much longer development cycles and b) quite different proof-points inherent in developing a product which must be medically certified.

Thus, we argue that in order to develop an AI in healthcare product, the entire development cycle must be changed. We present a rethought product development cycle, primarily aimed at products in the clinical setting, bridging the cognitive gap between modern digital development practices and current biotech practices. The purpose of this map is, on one hand, to enable potential digital health founders in the area of AI in healthcare to have a more accurate perspective of the road ahead, and to allow existing digital health entrepreneurs to better allocate their resources. On the other hand, the framework allows scientists, regulators, investors and political actors to better understand the specific needs of AI in healthcare projects and enterprises.

## Results

We have developed a framework following three consecutive phases and covering 4 domains for AI/ML-driven clinical product development. The 3 phases follow an entrepreneurial timeline. We break the process down into three distinct product development phases before a product exists in the market. Each of these phases is self-contained and typically can be recognised by both investors and regulators as a distinct stage on the road to market. For each of the phases, our framework spans across four major domains which are relevant for the development of an AI/ML product in healthcare: clinical, regulatory, data, and model development. Finally, we present an overview of the most important post-market risks, delivering software updates, changes in medical practice and surveillance, each of which must be mitigated against once the product enters the market.

In the following, we will give definitions of the 3 phases and 4 domains before outlining the framework in detail. The phases and domains are additionally depicted in Figures 1 to 4 following a domain perspective.

Definition of the 3 phases

Phase 1, called *Form*, involves bringing together a small team/group around a potential clinical need, with the purpose of providing a preliminary technical solution. This phase may take place within the support structures of a hospital spin-out unit or company incubator. In this phase, the main goals are to form, and maintain, the group membership, check the technical feasibility of a solution, and to understand the pathway to market by validating the clinical need. This latter point, we will explain, actually already requires a basic understanding of regulatory and clinical validation paths. At the end of the *Form* phase, the intrinsic value will typically amount to a proof-of-concept (PoC) solution and little more.

Phase 2, we call the *Build* phase. At this point, a more defined team decides to work together for a period ranging from 18 months to 5 years. Clinical members may initially maintain other commitments, but they need to provide enough availability to help steer the development process. The goal of this phase is to build a basic implementation appropriate from a clinical and regulatory point-of-view. The outcomes of this phase are multimodal, but the intrinsic value at its end-point is a solid codebase which demonstrably solves a clinical need.

Phase 3, *Launch*, is the phase which least matches current (consumer-oriented) software development practices. A medical product cannot directly enter the market without first passing certification and clinical validation steps. Therefore, processes such as medical device certification or a clinical study demonstrating efficacy must be carried out. In order to carry-out certification, or a study, typically the software version must be frozen. This does not mean, however, that the team stops working. In our framework, there is still considerable technical work in this period. That said, the main goal of the phase is to complete the certification and validation steps. The end result of this phase is a product which can be deployed, in the majority of appropriate clinical settings, in the market.

Definition of the 4 domains

*Clinical domain*

The clinical domain refers to the process of identifying real-world clinical needs and validating these needs throughout the life cycle of the project. Clinical needs are not just medical needs but extend to identifying different interaction requirements of different types of users (UI/UX research), and process-needs of all clinical partners, including patients and payers. We visualise, including supplementary advice, this domain in Figure 1.

*Regulatory domain*

Medical device regulation encompasses all laws and rules with regards to the development of a healthcare product falling under the definition of a medical device. AI in healthcare products are highly likely to fall under such regulations. We will explore for all three phases of a healthcare AI project, how to always be prepared for the regulatory requirements of the next phase and to deleverage critical risks early in the project. The particulars of regulatory processes differ enormously throughout the world, we will give only general advice which should apply across regions. A visualisation for the Regulatory domain appears in Figure 2.

*Data domain*

Since AI methods are inherently data-driven, the right kind of data is crucial to develop an AI-based healthcare product. In the data domain sections, we will explore which kind of data needs to be leveraged in each phase. The data necessary for an early proof-of-concept to get traction can differ tremendously from the data which is necessary in later stages for certification and for clinical validation. We will highlight for each of the 3 phases, what data sources founders and developers should focus on. Figure 3 contains an overview.

*Algorithm domain*

Another crucial domain is the choice of the AI methods applied for product development. With regards to the algorithms, attention must be paid: to the number of addressable patients in both the early training phases, but also in subsequent deployment; to questions of engineering around energy use, timeliness, and location of the data; to neglected populations in biased training samples; and, to the use of algorithms which will support any planned subsequent integration of other data streams. At every stage, we try to simplify this process outlining the long- and short-term priorities and how the two should align. Figure 4 outlines the algorithm domain results.

## Form Phase

The *Form* phase of a medical AI project can last 6-12 months and involves the earliest stages of proofing a concept around a digital solution to a perceived clinical need.

Clinical domain

The most important goals of the *Form* phase are to define a clinical need and to show that it is solvable in the form of a proof-of-concept. Clinicians are expert practitioners, and as such rely on their expertise to identify those clinical needs. Decades of experience in product design and development have shown, however, that while expert vision is useful for the identification of product-market fit, experts might fall too quickly in love with their own solutions and their own pet pain points [34,35]. This means, that the correct - objective - identification and validation of a clinical need is a high priority for the *Form* phase of development. The biggest risk for the *Form* phase is to focus on work that has no validated clinical need.

A clinical need is not only defined as a clinical problem that needs solving. Since development is the goal, the solution for the clinical need at hand must either provide a proven clinical benefit - thus doctors will be led to use it despite potentially higher treatment costs - or it must reduce costs. If the solution neither reduces costs nor shows a proven clinical benefit, then the potential product is not financially viable. The team that needs to come together in the *Form* phase will thus need to be able to identify the clinical need, to validate it and build a first technical solution to address this need.

Regulatory domain

Regulatory requirements are a significant hidden cost of development of medical AI solutions. Medical device regulators, such as the Federal Drug Administration (FDA) in the USA and European Medicines Agency (EMA) in the EU, have a mandate to protect the public from potentially dangerous medical interventions. This means that the choice to be regulated is rarely, in fact, a choice. Given this situation, entrepreneurs need to understand very early how their future product will be regulated. In the *Form* phase, it is unlikely that software developers will need to follow best practices for the development of a medical device. Most projects fall apart long before even a pilot study, so spending time and effort on regulatory compliant software-development processes at this early stage is inefficient. This is the time, however, for the founding team to familiarise themselves with the regulatory road ahead and to be aware of future requirements. Here, it is prudent to seek professional help as regulatory requirements are legally complex. This knowledge will build confidence with potential funders and will deleverage a large portion of *Build* (phase 2) risk ahead of time.

Data domain

New projects frequently lack access to fitting data or realise (too) late that access to promised data cannot be granted. Many clinicians are not aware of legal impediments to share data, for commercial use, so the decision to form a team should be based on clear evidence that data will be available and usable.

When such initial data is successfully secured it is paramount to evaluate whether the data is supportive towards solving the identified clinical need. Here, clinical founders often overestimate the value of their initial dataset. While it is certainly admirable, and indeed necessary, to bring data to the table at the outset of a medical AI project, the iterative procedure of product development will often make the utility of the initial data set redundant.

Any initial data set should be seen as an early research tool, from which to draw insights, which will contribute to the product development process. In the *Form* phase, it is enough to build models which can only predict outcomes based on this, typically, single-sourced - with multiple hidden biases - data. Practitioners should, however, begin to test the limits of their assumptions about this data. The urge to cherry-pick and to oversell the data utility, at this point, is only storing up larger problems for later. Be aware that a thorough validation of data hypotheses will be unavoidable in later stages. It is thus crucial to come up with a data plan outlining the usefulness of the currently used data and defining the need for future data sets.

Algorithm domain

Similar to the data domain, models developed in the Form phase will often be super-seeded in later stages. What is important in this early phase is to choose algorithms which are fit for purpose, not just with this smaller data set, but also with data on the intended deployment scale. Testing of multiple algorithms and the development of performance benchmarks are also very important. Choosing methods which over-perform on the initial data set, but which store up problems for subsequent expansion, along with performance metrics which are inappropriate for clinical applications are common errors at this point.

A tempting self-deception is to fall into the habit of adopting bad scientific practices to produce initial results which in turn lead to very high chances of failure in later stages. We understand the need to show quickly initial results which will be the basis to attract further funding - be it public or private. However, sloppy development in this stage will lead to high failure rates in the *Build* phase, even if a clinical need exists and the team has the right data to solve it.

Outlook

*Form* is a crucial phase where a few vital key objectives need to be kept in mind. Foremost, the focus must lie on the validation of a clinical need and the development of a proof-of-concept to solve it. The clinical need must lead to reimbursement - either through proven clinical benefit or proven cost reduction. The formalisation of the data and algorithmic hypotheses, and associated benchmarks, is a demanding process for smaller teams. Thus, it is essential to focus on the key objectives.

## Build Phase

Following the early incubation of the *Form* phase, comes *Build,* which can take as long as 2-5 years depending on the type of product being developed. In this phase, a concerted effort is made to develop an AI-driven solution, while also demonstrating a clear clinical need.

Clinical domain

During the *Build* phase, it is highly important to continuously validate and fine-tune the clinical need, to make sure that the development process follows the right path. Due to the multi-modal nature of product development, a typical product will morph multiple times during this phase. A strong risk, in this phase, is that one ends up with a technical solution which works well but no longer addresses the actual clinical need. Another potential risk is that the solution works on the available data but is actually non-performant on data from other sources (i.e. a highly biased model is developed). Thus, it is important to work with partner clinics, or practitioners, which are not directly under the control of the project initiator. This is partly to avoid complications of medical hierarchy, but also to expose the product development team to differences in clinical practices, hardware and opinions. Only this exposure will allow the team to fine-tune the solution to real-world needs. Especially in the Build phase, it is crucial to follow a predefined and continuously updated clinical validation plan.

Regulatory domain

The *Build* phase involves an earnest attempt to develop software which will form the bedrock of a future product. As such, this is the kick-off point for development of a solution which will withstand regulatory approval. There is the evident risk that incorrect assumptions are made about the required certification levels and the associated costs and timelines. In the worst case, this discovery is made so late in the *Build* phase that mistakes cannot be undone and the project or startup fails. It is also important to make sure that every single piece of technology is used in the correct manner for the certification process.

There is a strong trade-off between instituting correct medical device (MD) certifiable software development processes early vs late in the product development cycle. The process overhead for developing an MD certified product is considerable, especially in a product which is not yet tightly designed to fit a product-market niche. However, introducing retrospective audits and new development processes late in the development cycle can be extremely financially costly and time-consuming. Potentially tipping the balance in favour of introducing MD certifiable guidelines early in the process, is the fact that this will also bed-down appropriate practices in the development team. Since these practices will need to be continued for the entire lifecycle of the product, it makes sense to begin as early as feasible. It is clear, however, that for some products, a later start might be the better option. The key is that decision-makers actively make an informed decision as to how to proceed and when development will switch to the scrutiny required for the certification process.

Data domain

This is the stage at which external data sources must be aggregated with internal ones, allowing hypotheses about homogeneity and data quality to be assessed systematically for the first time. This, however, requires a strong focus on data processing and harmonization. It is important to make sure that the team has not only machine learning expertise but is also experienced in correct data processing. Since any work carried out in the *Build* phase is typically in very close partnership with clinics, or partners, it allows a relatively tight feedback loop on data which will not exist in later stages of development. Here, new datasets allow continuous assessment of the model performance. It is crucial to emulate in the *Build* phase the heterogeneity of the real world, e.g. in terms of different hardware or different clinical practices.

Depending on the clinical need, it might be necessary to run a data collection study in the *Build* phase. For example, data from a new study might be necessary to reflect a change in current clinical practice, where retrospective data might not exactly reflect the use case anymore. It is also possible that in the *Form* phase the proof-of-concept was purely technical and data for the specific use case was not, at that point, available. It is crucial to understand that the data set, on which the medical AI product is built, can rarely be used for the subsequent validation of the product itself (see *Launch* phase and Discussion). At the end of the *Build* phase, the dataset must one-to-one reflect the real-world conditions to allow the development of models which will perform as well during medical certification and clinical validation as in-house.

Algorithm domain

Model development in the *Build* phase is tightly coupled to issues with data and clinical fit. With access to a larger data set, priority should be set on maintaining the predictive value, despite potential differences between the sources. It is not uncommon that the initial technology is unsuited to accommodating heterogeneous data. Thus, it happens frequently that the models need to be rebuilt from scratch in this phase, often with the assistance of more experienced engineers.

Associated with predictive value, the single biggest issue in the *Build* phase is the mitigation of bias. The developed models must maintain predictive value on all real-world data which the future product will encounter and not just on a subpopulation. Patients from different regions can exhibit very different disease patterns, for example. In image generating fields, like radiology/pathology, different hardware (scanners/digitalization software) and different clinical practices (sequence settings/staining practices) can lead to tremendous differences in image statistics. Thus, funders and founders should not be blinded by high predictive values on single-source data. Models developed under such conditions can fail miserably when confronted with even slightly different datasets.

From an engineering point of view, the finished product will be deployed somewhere and must fit into the clinical workflow. Choosing algorithms which both scale appropriately and are deployable in the context is vital. Algorithm designers should also bear in mind extensibility, a coherent product roadmap might incorporate plans for future data streams enabling potential extended applications.

The *Build* phase is also the phase where the necessity of explainable AI (xAI) deserves to be considered. Applications, such as clinical decision support, might require explanations as the users (doctors) demand them [36] or are required to use them[37]. A more comprehensive product vision might even leverage xAI to provide different levels of explanation to different classes of users (e.g. doctors vs nurses vs patients). For the developers, it might be prudent to utilise xAI to identify potential hidden prediction biases. For example, it could be that a certain pathology is coupled to a certain hardware characteristic in the data. In that case, the model might learn to identify that hardware-tag instead of the pathology [38]. Most importantly, there are signs that xAI might be required by medical device regulatory authorities in the future [26].

Outlook
*Build* is the phase which most closely corresponds to the public perception of product development. Here, clinical validation should ideally switch from identifying the clinical need to fulfilling it. It is also the phase in which most companies stagnate. Reasons for that can be discoveries which indicate that initial assumptions about data were not accurate. Software solutions may be found to contain bugs or are more difficult to build than early results suggested. The largest effort, in this phase, must be applied to bringing all the components together. This applies on the software side, requiring not just the development of an AI tool but also a practical user interface (UI). But it applies equally to the validation side, users can never be forced to use your tool, they need to want to use it. Getting to this point requires many cycles of user interviews and testing. In *Build*, it is still possible that the true nature of the clinical need has been misinterpreted by the development team, or that the proposed technical solution is not capable of attaining the required level of performance. Ideally, however, this phase ends with a product which is ready for medical device certification and efficacy studies and all major risks for the *Launch* phase have been kept as low as possible.

## Launch Phase

The last phase, *Launch*, can take up to 3 years depending on the type of medical AI product. In this phase medical device certification is achieved and clinical validation is performed.

Clinical domain
The final clinical validation is proof that the product fulfils its promise in the clinical real-world setting. There are a few cases where retrospective data (also referred to as real-world evidence) might be used to 'prove' clinical products. There are clinical, regulatory and even statistical reasons [39] to argue that these cases will always prove to be a very small minority (see also the Discussion). In general, then, it will typically be necessary to run a clinical trial of any new AI-based medical product. The goal of this trial will be to prove not just safety but also clinical efficacy. This step is crucial. Without proven clinical efficacy - either for clinical benefit or for cost reduction - there is no real incentive for customers to buy the product. Thus, the successful completion of such a trial is usually the final step before widespread sales of the product in the medical-product marketplace can happen.

It is crucial to understand that such a trial will almost always have to be a randomized controlled trial (RCT). Only RCTs can provide the necessary evidence level which is required to change clinical practice and to find their way into clinical standard-of-care guidelines. RCTs can also be designed to allow identification of effects on sub-populations, which is often necessary for AI-based products. A major risk for the *Launch* phase is to underestimate the time and costs required for the planning and conduct of a RCT trial, which can amount to several years.

Any evidence established by the trial will be subsequently examined under the lens of the trial design. A trial closer to real-world conditions will bring a higher value to the product (and will be necessary if the route to market requires changes to standards of care), but it runs a higher risk of failure due to uncontrolled conditions. This risk is mitigated by practices in the previous phases working with heterogeneous datasets and minimizing bias. To ensure an optimal trial outcome, it is highly advisable to allocate necessary funding and to work as early as possible with a contract research organisation (CRO). Particular attention should be paid as to whether clinicians, in the trial, actually use the software in the manner expected by the designers. Once the product is approved and it appears on the market, further validation and changes will be extremely costly.

Regulatory domain

In many senses, the *Launch* phase is the regulatory phase. In it, the team will focus on the implementation of a fully compliant software development process and all steps necessary for medical certification will be taken. The major regulatory risk for this phase is that key assumptions about the certification process were wrong and the certification process is massively delayed. Importantly, this phase will also involve considerable efforts to consider the regulatory aspects of a product post-launch, when it is already in the market. Processes must be developed to handle surveillance of how the product behaves in real-world usage. If users report a bug, which may cause harm to life (i.e. an adverse event), product managers must already have contingency plans in place as to how this situation will be handled. Given the current state of medical device certification, negotiations must also be entered into with regulatory bodies as to how minor and bug-fixing software updates will be handled without requiring a re-run of clinical trials. A software solution which requires a complete clinical trial for every minor bug fix will not be worth much. By addressing these issues before they arise regulators will be more willing to put in place framework agreements laying out conditions for automatic certification of updates.

Data domain

The *Launch* phase is, in most cases, the final phase in which the development team can expect to see new forms of medical data. For certain cases, clinical trials can be constructed such that the internal team are, subsequently, allowed to train their models on the newly acquired data. Such a process will likely require backtesting, demonstrating that, had the new model been used throughout the trial then the algorithm behaviour would have been identical. What happens, however, in the case where backtesting indicates minor differences in treatments is still an open debate in regulatory circles. Notwithstanding, a major risk is that the data used in the previous stages were not heterogeneous enough and real-world data within the clinical validation is much noisier or so different that the predictive value does not hold. It is also paramount to ensure that the data collected in the clinical validation study actually represents the real-world setting. Here, it will be necessary to obtain proof for this claim. It should be kept in mind that the acquisition of patient data directly from post-launch usage of the product is likely to be rare. It is true that a product which, through its usage, accretes additional knowledge about patients is an incredibly valuable product design. However, patient privacy practices throughout the world vary to such an enormous degree, that we find it unlikely that many products will successfully follow this path. Thus, data collection in this final pre-market phase must be planned and performed with the highest scrutiny.

Algorithm domain

The greatest risk in the *Launch* phase is that the final software solution will fail. It should be mentioned that at this stage, next to model failure, design failure can also be the cause. Extensive testing in *Build*, both on the model development and the UI/UX side, should help to keep the risk of such an outcome as low as possible. That said, AI models and software solutions around them fail at all stages of development. Typically, in *Launch*, this will occur due to some previously unrecognised regularities in in-house data ('house effect') which does not transfer to real-world data. Assuming such a catastrophic failure does not occur, algorithmic focus should be on how well the model transfers to out-of-house data. Often, despite all previous efforts, real-world data is still noisier and less homogenous than previously acquired data. The *Launch* phase is the last phase in which it is possible to easily detect lower performance on lower quality data and make final adjustments to the algorithms. In order to build a product which will reach market acceptance, it is important to be prepared for this challenge, to embrace it and to solve it. The team should be experienced enough by now.

Outlook

For a general non-medical digital start-up, the third phase is the one in which advice turns towards rapidly expanding the user base and optimising the product-market fit. In a medically regulated product, however, the third phase must precede rapid growth since the product needs to become certified and clinically validated. The processes for certification and clinical validation remain, to a certain degree, very much the last chance to get product-market fit aligned. Minimizing the risks in the previous phases of development and working with professional partners both for regulation and clinical validation will keep the risks for failure low in this final phase.

## Post-market

After completing *Launch,* the product development roadmap returns to that more typically seen in non-regulated markets. A certified medical AI product has been shown to be safe. If it has been through an RCT it may also demonstrate certain standards of clinical efficacy. In most cases, this means that a sales process can proceed and the delivery of the product massively scaled-up. Addition of new product features will usually follow a slightly condensed version of the *Form-Build-Launch* cycle. For the product in the market, three topics remain of considerable ongoing concern: software updates, data practices and surveillance.

Software, as it is developed today, is incredibly buggy. Even when following advanced software development best practices (e.g. B-Method [40]), software errors persist and must be corrected for. In physical medical devices, the electronic solutions follow decades of engineering-led best practices. In the event of a problem, the scale of the problem (risk) is analysed and regulators usually influence the decision as to whether to recall the product or merely fix it in future models. Software combines the ability to update the working system remotely, at almost zero cost, with a lower quality of initial engineering. Since software for medical purposes has seen relatively little development to-date no clear working regulatory practice has yet emerged. What is clear is that medical products are evaluated in terms of risk: What risks does this product pose to the patient? What if it goes wrong? What must happen if it goes wrong? This evaluation is carried out at the macro-level of the whole product, at the micro-level of the individual components, and at the intermediate-level of the interface which brings the components together. Development teams need to stay on top of shifting standards of best practice and, in all likelihood, should consider medical AI development to require similar levels of oversight as that historically required by aeronautical, spaceflight and automobile software development.

A further, often overlooked process, in developing a medical AI solution is the question of shifting clinical standards. AI experts refer to this problem as *stationarity of the data*. For example, the diagnostic criteria, and how they are encoded in medical records, change over time. This is an expert-driven process which trickles down through medical practitioners. Development of a medical AI usually happens in a concentrated window in time, during which medical practices might be relatively stable, but the deployment of such a solution must then be used over a much longer time period. Detection, whether by automated means or by internal human-driven processes, of changes in medical practice must be planned-for and processes established as to how to update the software in response. This is a huge challenge, as the whole process of *Form*, *Build* and *Launch* might be based on a single snapshot of clinical practice. A major shift in clinical practice can invalidate most of the work done during development.

Surveillance is a concept which healthcare professionals and pharmaceutical manufacturers understand reasonably well. It is not something which software and AI developers will naturally consider without prompting. There must be feedback methods for reporting misapplications of the software in the wild. The spectrum of events which must be accounted for ranges from actual medical adverse events to users accidentally misunderstanding certain display elements. Most likely these methods must span the spectrum from paper-based reporting methods to in-device automated monitoring. Furthermore, awareness must be paid to the fact that streaming of telemetry is usually frowned upon in medical circles; it is too easy to leak patients' private medical data. We are not aware of good examples of medical AI device surveillance solutions which do not require ongoing access to patient medical records and extensive use of clinical review panels. No matter what the solution arrived at, it is the company producing the medical AI device which will be responsible for product surveillance and must pay for ongoing support in this domain or face a loss of medical certification.

# Discussion

Healthcare systems all over the world face the same challenges. There is an urgent need to deliver better quality healthcare and simultaneously lower costs to keep our healthcare systems sustainable. Artificial Intelligence has the potential to fulfil both of these requirements. While AI often hits the headlines with high performance in concept studies, certified and clinically validated products are still rare. To battle this *translational gap*, we present a framework addressing best practices for the development of AI healthcare products.

The presented framework is an original and unique analysis of the development of AI healthcare products. It divides the development into three consecutive phases: *Form*, *Build* and *Launch*. Each phase is set to fulfil important requirements and deleverage the major risks of the subsequent phases. These three phases represent distinct stages as they appear in the course of the development of AI in healthcare products. Researchers, founders and other stakeholders should be able, with the aid of our framework, to easily identify the current phase and relative progress of a project. Within these three phases, the framework covers the 4 most important domains of such a project, namely the clinical, regulatory, algorithmic and data domains. Each domain must meet strict requirements in order to successfully complete the phase. If one of the areas is lagging behind, this will lead to the accumulation of - potentially fatal - risks for the project and jeopardise the product.

We are not aware of other frameworks focusing on AI in healthcare product development to date. Our approach largely follows the "decision perspective" model introduced by Krishnan and Ulrich [30]. Its advantage is that it does not rely on the perspective of a single domain, but rather divides the development process in a series of decisions. Thus, for each phase, the right decisions can be made to achieve specific goals. This, of course, paves the way for future research work, which can focus on more specialised areas of AI in healthcare development such as marketing, organisations research and other topics. Importantly, the advantage of a decision perspective framework is the focus on *what* to do and not *how* to do it. Thus, frameworks focusing on how to best perform development are complementary to our framework as well as field-specific roadmaps aimed at the individual challenges to adopting AI [41]. The literature about this topic is too broad to be covered fully here. We will give a short overview of some frameworks in use. A key discovery, of the past 20 years, was the coining of the concept of an 'effectual entrepreneur' [42]. Such an entrepreneur works at the extreme end of low inherent knowledge - unplannable processes. Over time, they develop an enterprise which increases inherent knowledge and beds down processes, which ultimately lead to a product. In start-up culture, such an entrepreneur is characterised by using product life cycle approaches such as Lean Startup or Agile (not to be confused with Toyota's Lean manufacturing) [32,43]. Larger industries, attempting to learn from start-ups have adopted Design Thinking [44] - an essentially two-step process which begins by focusing on need before moving to Lean Startup-style product development. From a funding point-of-view, the Business Model Canvas [45] has proved invaluable in identifying the product-market fit for new companies. Finally, the Three Lenses of Innovation -  a potential product must be *desirable*, *feasible* and *viable* - has proven useful to examine prospective market needs and product-market fit [46]. Our framework is both compatible with and extends these frameworks.

Importantly, within the framework, we deliberately did not focus on funding schemes. The reason here is that current funding schemes often lead to a misalignment of interests. The reasons for this status-quo are complex. Current funding schemes for AI in healthcare products - often termed Digital Health - usually follow experiences from products outside of healthcare, derived in the modern digital era, where a sufficient return of investment (ROI) is expected within a few years. However, despite being digital software products, AI in healthcare products are much closer related to biotech products sharing the complex development and heavy regulatory requirements. Why then are funding schemes from biotech not adapted to the AI in healthcare sector? This can be attributed, in part, to the fact that the main exit for biotech companies is a combination of initial-public-offering (IPO) and merger and acquisition (M&A) [47], usually in partnership with large biotech or pharma companies. In general, outside of biotech, IPO activity has significantly decreased [48], meanwhile few large acquiring companies have emerged for medical AI. Thus, it is very hard for investors to ensure that a potentially very large investment sum will yield the required ROI. Relatively few investors are even aware of this challenge. Founders, whether unscrupulous or naive, attract investment by advertising their product niche as being equal to any other (AI) software. Or, by categorizing the product as a non-medical device, with stark consequences in terms of limiting their abilities or facing a regulatory backlash. Thus, AI in healthcare development is forced into following development paths of simpler products due to unreasonable expectations. This inevitably leads to founders cutting corners to pretend fast agile development. Evidently, such endeavours are bound to fail, when the hard requirements of medical device regulation and clinical validation cannot be met. In the words of venture capitalists, the main challenge for AI in healthcare startups is the so-called "valley of death", where the startup already exists but does not generate revenue. This phase is usually equivalent to the *Build + Launch* phases in our framework. Thus, there is a need for an honest appraisal of the development roadmap for AI healthcare products. Here, our framework is an important contribution. Early-stage funding needs to be made aware of the true roadmap and timelines, and expectations need adjustment accordingly. As an example of increased public funding, a scheme specifically targeting the valley of death is currently set up by the European Union launching in 2021 [49]. As the sector matures, we hope that there will be an expansion of the acquisition activities by large pharma- and med-tech companies leading to a clearer pathway to product and company development.

The above-identified crisis of the funding models for AI in healthcare products opens up the question as to how AI in healthcare products will be funded in the future. The current industry response which we have identified is to lobby for reduced regulations. The proposed rationale for this is that faster and less regulated technological innovation will ultimately deliver on the healthcare goals of broadening access while reducing costs. This cost argument has led to a willingness, on the part of regulators, to open up avenues for clearance of medical software solutions without the full burden associated with traditional pharmaceutical drug and medical device regulations. This opening of alternative regulatory routes is tempered by an awareness that medical AI can, and will, directly impact on patient health. In practice, however, the reduction of regulatory barriers appears to lead to a potential diversion of spending, from clinical validation and software engineering to paying experts on the navigation of such shortcuts - and a subsequent enormous payment for product-risk insurance when the product reaches the market. We do not see this as a sustainable path, indeed much of the current excitement around healthtech fails to appreciate that sales in this space are typically subject to subsequent regulatory approval. Additionally, the ethical implications of potentially endangering patients' health in a trade-off for faster development are tremendous. Instead of reducing regulations, we suggest a more mature approach to clinical validation based on realistic approaches such as that contained in our framework. That said, there are alternatives. We see huge scope for the certification of AI 'platforms' or manufacturers where only the new components or the manufacturer need to be certified upon upgrades [50].

A related narrative is the increasing discussion of Real-World Data (RWD) as the real goldmine for AI in medicine. Currently, efforts are being made to allow Real-World Evidence (RWE) for the clinical validation of healthcare products [39,41,51]. What is the challenge of RWD for clinical validation with regards to AI in healthcare products? Importantly, the term 'AI' only describes the method used in the product, not the model of medicine which lies behind it. Here, a significant observation is that many AI-solutions developed and marketed today can be considered - often unbeknownst to their creators - precision medicine approaches [52]. Precision medicine is a type of personalized medicine aiming to integrate all available data for each patient into an individual profile and tailoring diagnostics or therapy to this profile. AI is a great fit for precision medicine due to its inherent ability to utilize high-dimensional data to identify subpopulations and make individualized predictions [53]. Thus, RWD often allows the development of AI products - in the *Form* and *Build* phases - as, currently, only RWD has available data for sub-populations. In this sense, RWD can be seen as a driver of AI development. Data from classic randomized clinical trials (RCT), typically lack the granularity to distinguish differing populations and have - often - less value for AI in healthcare products. RWD is per definition retrospective and thus has - all widely known and accepted - drawbacks, e.g. bias, lack of controls, confounding and others [54]. Here, the main challenge is that availability is not a valid surrogate for the required level of evidence for safety and efficacy. One of the greatest hidden risks in a medical AI product developed on RWD is in algorithmically enforcing bias. And indeed, commercial products in healthcare have already been shown to contain considerable racial bias [55]. RCTs, in contrast, have been developed to minimize bias and confounding and provide the highest level of evidence. It is perfectly possible to perform an RCT to show efficacy and safety also for subpopulations and hence for any AI in healthcare product. The main difference is that a subpopulation RCT is very costly and time-consuming, as a sufficient number of patients per sub-group need to be enrolled. A push to use RWE instead of RCTs for the validation of AI in healthcare products is essentially a push to trade the level of evidence of safety and efficacy for lower costs, increased speed of development and higher risk for patients due to bias and confounding. It is fair to say that such a push would ethically and publicly never be acceptable if it was suggested for pharmaceutical products. Thus, it is hard to justify such a move for AI in healthcare products which often might have the same potential for patient risk. Consequently, our framework stresses the importance of an RCT as the required level of clinical validation for the overwhelming majority of AI in healthcare products in the *Launch* phase.

In developing our framework we have largely envisaged medium- to high-risk medical AI products. This has streamlined our thought processes and allowed us to develop rules particularly of use to a medical audience. However, we don't see any real differences in the process for developing low-risk products. The only place where there is a question mark is in products which attempt to fall below all risk thresholds for medical regulation. At this point, we would question the wisdom of any strategy which looks at providing clinically-oriented benefits while attempting to market itself as a non-medical product. This continues to run the risk of falling foul of regulatory attention while also failing to attain the enhanced sales-values associated with medical products. In any case, a familiarity with our framework will allow key decision-makers to evaluate the potential costs and benefits of following a regulatory path.

## Conclusion

We have developed a framework which outlines, what we consider to be, best practices in AI in healthcare development. Our goal is to pave the way for products which are both efficacious and safe for patients. Through direct participation in the development of a spectrum of medical AI products our experience, and that of our peers, is of enormous waste in the process. Current initiatives to improve the flow of AI in healthcare products aim at reducing costs and speeding up development through means which potentially increase risks for patients. Our framework outlines an alternative to this approach. We aim to improve the development situation of AI-driven medical products by enhancing accountability, transparency and planning, hopefully increasing success rates for such products, without facing some of the safety and ethical trade-offs mentioned above, to fulfil the promise of AI in healthcare: better care at lower costs.

## Disclosures

## Acknowledgements

We would like to acknowledge Michelle Livne, Nicholas Borsotto, Tim Higgins and Dorothée Doepfer for valuable ideas and proofreading.

# Clinical Validation

|  | Form | Build | Launch |
|---|---|---|---|
| **Risks** | - There is no clinical need<br>- False assumptions about the clinical need<br>- Inability/failure to form a team which can solve the clinical need | - Inability to show clinical benefit<br>- PoC product is not robust to clinical conditions<br>- A mismatch between pilot product and actual clinical need | - Clinically validated performance is considerably lower than pilot studies<br>- Users don't understand or by-pass the solution<br>- Time and cost of clinical validation are underestimated |
| **Objectives** | - Demonstrate clear evidence for a clinical need<br>- Demonstrate that the clinical need is technically solvable (Proof of concept) | - Demonstrate a clinical benefit in several pilot environments<br>- Evaluate the level of (one-off) customisation needed in delivering a solution to the clinic<br>- Demonstrate that users are actually willing to use the product<br>- Discover potential clinical barriers to deployment | - Clinically validate the efficacy of proposed product<br>- Ensure that clinical validation resembles the actual clinical setting<br>- Demonstrate clinical safety of product<br>- Evaluate degree to which 'naive' clinical users are willing to engage with product<br>- Develop a system for monitoring changes in clinical practice |
| **Key Results** | - Basic market research / interviews completed indicating a real clinical need<br>- Promising initial results, derived from a robust approach, suggesting the clinical need can be solved (see also data section)<br>- Proof of concept | - Deployment of prototype systems working in several clinical environments<br>- Evidence that clinical need is solved across patient populations<br>- Interest from clinical partners to continue using the software | - Extensive UI/UX testing<br>- Third-party validation of product in a non-controlled environment (usually RCT)<br>- Procedures for detection / tracking of updated clinical standards |
| **Advice** | - Develop market-fit / clinical-need hypotheses<br>- Test these hypotheses early<br>- Cycle through them on an ongoing basis<br>- Test the bigger (cost weighted) risks first!<br>- Demonstrate clearly proven clinical benefits (doctors are forced(!) to use the solution) or reduce costs. If you will do neither, nobody will pay for your solution. | - Observe how technically naive users interact with pilot product<br>- Do not rely on testimonials<br>- Gather use statistics<br>- Be open to surprising results of your validation that might change your product direction<br>- Make a validation plan, iterative over it continuously | - Be aware that you cannot sell a product into a non pilot study environment unless it is demonstrably safe<br>- Users will avoid change unless you provide clear, visible, proven benefits to them<br>- Choose the trial with smallest efforts and lowest costs. It might be beneficial to work with a contract research organization. |

# Regulatory Affairs

|  | Form | Build | Launch |
|---|---|---|---|
| **Risks** | - Lack of awareness of future regulatory demands which will impact on development methodology | - Wrong assumptions about required certification levels and associated costs and timelines<br>- Late discovery that key technology was not developed in a medically certifiable manner<br>- Not using certifiable development methods now for a piece of technology which will ultimately be a part of the product | - Key assumptions about the certification process are wrong and delay certification massively<br>- Lack of preparation for the post-launch phase lead to a "frozen" product which cannot be deployed |
| **Objectives** | - Understand regulatory demands for the Build phase<br>- Develop initial regulatory strategy | - Demonstrably follow best practices for development of 'safe' (ie. medically certifiable) software and devices | - Fully documented, certification compliant software development process<br>- Develop product surveillance plan<br>- Develop product update procedures |
| **Key Results** | - Initial regulatory strategy | - Clear process pathway to certification, or initial certification<br>- Internal processes which are appropriate for selected level of medical certification | - Medical product certification<br>- Procedures for deployment of software updates<br>- Procedures for surveillance and reporting |
| **Advice** | - Seek professional legal assessment of the regulatory requirements for possible future product | - Understand and implement the correct Medical software development process<br>- A key decision for your product is to know when to start with the implementation of medical device development tailored to medical device regulations.<br>- UI/UX designed and tested to prevent overstatement of / misleading results<br>- Contact appropriate regulatory institutions early<br>- Work together with professional companies specialised on medical device certification<br>- Understand the specific AI-focused regulatory requirements of your country/region<br>- Understand the certification process of your region fully.<br>- Make contact to regulatory bodies early in the process.<br>- Work with a partner specialized on medical regulatory affairs | - Do not underestimate the complexity and burden of the final implementation of regulatory requirements<br>- Consider automated detection of non-handleable cases<br>- Work with a partner specialized on medical regulatory affairs |

# Data Strategy

|  | Form | Build | Launch |
|---|---|---|---|
| **Risks** | - Lack of access to data<br>- Not enough data/signal in data to develop proof-of-concept<br>- Assumptions about the statistical characteristics of data are wrong<br>- Lack of understanding what kind of data is needed for later stages (beyond proof-of-concept) | - Site-to-site variability in clinical data is greater than expected (data harmonization fails)<br>- Increased access to data does not lead to hoped for improvements in signal detection<br>- Failure to acquire extended data access rights | - Real-world data has lower fidelity than that obtained from prototyping sites<br>- Real-world patients are less homogenous than those in early data sets |
| **Objectives** | - Acquire access rights for initial data set<br>- Understand the structure of the clinical data (width (D), depth (N), regularity, signal/noise, bias)<br>- Determine appropriateness of data set to solve clinical need<br>- Construct a data plan for data you need in later stages | - Data harmonization of different data sources<br>- Obtain legal access to multiple new datasets | - Ensure that data collected in the clinical validations is equivalent to the real clinical setting<br>- Determine whether the general (clinical) population have the same data characteristics as those used to build the algorithms |
| **Key Results** | - Fitting data source identified<br>- Data is acquired / collected<br>- Data structure is understood<br>- Hypotheses formed, and validated, about ability to solve clinical need<br>- Data plan developed | - Sufficiently sized and diverse data has been obtained<br>- Harmonized database including all data sources | - Sufficient data with the pre-defined characteristics are gathered to allow clinical validation<br>- Data is similar to previous pilot studies |
| **Advice** | - Explore free datasets<br>- Partner with clinical institutions (study data or RWD)<br>- Utilize an interdisciplinary approach (understand what each feature and label exactly means)<br>- Be aware that you have little to no control over what is captured<br>- Focus in this stage should be on whether the insight actually makes sense<br>- It is appropriate to develop the PoC on potentially biased data, but be aware of it<br>- Use your PoC to determine how many data points you are likely to need for a product | - Avoid, identify and mitigate bias<br>- Look out for corner cases<br>- Make sure you have the data for all populations that are covered by your clinical need<br>- Hardware from different manufacturers produce different data sets. Make sure you have data from all.<br>- Remove locally specific identifiers. If not, your model might identify those instead of the pathology<br>- At every point of the process, make sure that predictive value holds up<br>- Make sure your team has enough people dedicated to data processing and harmonization | - Caveat: data heterogeneity can be a real problem<br>- Ask yourself this question often: How good is your system in the real world? |

# Model Development

|  | Form | Build | Launch |
|---|---|---|---|
| **Risks** | - Wrong methods used<br>- Wrong performance metrics applied<br>- Method appropriate for initial small dataset, but not for larger dataset (no scalability)<br>- Bad science | - Expanded access to training data does not lead to expected performance improvements<br>- The algorithm does not scale well<br>- The developed model is biased and does not reflect real-world conditions | - Model designed in pilot studies fails to work in real-world setting |
| **Objectives** | - Determine algorithmic targets (e.g. labels) which are in keeping with the clinical need<br>- Determine algorithm which is best suited to taking given data set and delivering on algorithmic targets<br>- Estimate algorithmic scalability for expected data sets<br>- Estimate algorithmic scalability for expected deployment scenario<br>- Determine early benchmarks for predictive value | - Test how well the algorithm performs in pilot environments and on heterogeneous datasets<br>- Identify<br>- Determine whether the solution technology scales well to new data and usability requirements<br>- Determine whether you need explainable AI in your product | - Deploy the algorithm at scale<br>- Determine how the solution copes with missing data, poor quality data and out-of-sample data<br>- Determine the predictive value of the solution in clinical validation. |
| **Key Results** | - Candidate best algorithm<br>- Estimates of predictive value on initial data set<br>- Estimation of how candidate algorithm will scale with future out-of-house data (bias assessment) | - Trained algorithm on expanded data set<br>- Predictive value holds on harmonized dataset<br>- Scalable architecture for deployment<br>- Tested deployment in different clinics | - Deployment in non-direct partner clinics<br>- Algorithm is robust versus missing data, noisy data and out-of-sample data<br>- Predictive value holds in clinical validation |
| **Advice** | - Apply the highest level of methodological AI scrutiny (no shortcuts)<br>- Utilize an interdisciplinary approach | - Make sure that your model does not rely on source specific hidden meta-data<br>- Latest in this stage you should know how to deal with missing data<br>- Recheck basic assumptions about data and your algorithms before introducing more advanced architectures<br>- Acquiring more data should (typically) have higher priority than advancing the architecture | - Approach should be based on level of risk.<br>- Consider rebuilding the software from scratch due to technical debt<br>- Make sure your development team is ready for post-market requirements |